\documentclass[nofootinbib, twocolumn,prl,aps,groupedaddress]{revtex4-1}
\usepackage{amsmath}
\usepackage{amsfonts}
\usepackage{amssymb}
\usepackage{graphicx}
\usepackage{bm}
\usepackage{color}
\usepackage{multirow}
\usepackage{lineno}
\usepackage{times}
\usepackage{dsfont}
\usepackage{siunitx} 
\usepackage{lipsum}

\usepackage{xcolor}
\usepackage[%
  colorlinks=true,
  urlcolor=blue,
  linkcolor=blue,
  citecolor=blue
]{hyperref}
\hypersetup{backref,
colorlinks=true,
linkcolor=blue,
linktoc=black,
citecolor=cyan,
urlcolor=black}

\newcommand{\kb}{\ensuremath{\text{k}_\text{B}}}

\newcolumntype{P}[1]{>{\centering\arraybackslash}p{#1}}
\newcommand{\pr}[1]{\ensuremath{\left[#1\right]}} 
\newcommand{\pc}[1]{\ensuremath{\left(#1\right)}} 
\newcommand{\px}[1]{\ensuremath{\left\lbrace#1\right\rbrace}} 
 
\newcommand{\ket}[1]{\ensuremath{\left\vert#1\right\rangle}} 
\newcommand{\md}[1]{\ensuremath{\left\vert#1\right\vert}}

\newcommand{\cRL}[1]{ \textcolor{black}{#1}}

\newcommand{\cccRL}[1]{ \textcolor{black}{#1}}

\newcommand{\bs}{\boldsymbol}

\definecolor{burgundy}{rgb}{0.5, 0.0, 0.13}
\definecolor{denim}{rgb}{0.08, 0.38, 0.74}
\definecolor{midnightgreen}{rgb}{0.0, 0.29, 0.33}
\definecolor{sienna}{rgb}{0.53, 0.18, 0.09}
\definecolor{sacramentostategreen}{rgb}{0.0, 0.34, 0.25}

\newcommand{\mi}{\mathrm{i}}

\DeclareSIUnit\gauss{G}

\begin{document}

\title{
Production and stabilization of a spin mixture of ultracold dipolar Bose gases 
}

\author{Maxime Lecomte}
\author{Alexandre Journeaux}
\author{Julie Veschambre}
\author{Jean Dalibard}
\author{Raphael Lopes}
\email{raphael.lopes@lkb.ens.fr}
\affiliation{Laboratoire Kastler Brossel, Coll\`ege de France, CNRS, ENS-Universit\'e PSL, Sorbonne Universit\'e, 11 Place Marcelin Berthelot, 75005 Paris, France}

\begin{abstract}
Mixtures of ultracold gases with long-range interactions are expected to open new avenues in the study of quantum matter. Natural candidates for this research are spin mixtures of atomic species with large magnetic moments. However, the lifetime of such assemblies can be strongly affected by the dipolar relaxation that occurs in spin-flip collisions. Here we present experimental results for a mixture composed of the two lowest Zeeman states of $^{162}$Dy atoms, that act as dark states with respect to a light-induced quadratic Zeeman effect. We show that, due to an interference phenomenon, the rate for such inelastic processes is dramatically reduced with respect to the Wigner threshold law. Additionally, we determine the scattering lengths characterizing the s-wave interaction between these states, providing all necessary data to predict the miscibility range of the mixture, depending on its dimensionality. 
 \end{abstract}
\maketitle



In the last decade, long-range interactions have become a central focus for quantum simulation \cite{Baranov2012, Bohn2017, Browaeys2020, Chomaz2023}. From arrays of Rydberg atoms trapped in optical tweezers to the first observation of a dipolar molecular Bose-Einstein condensate (BEC) \cite{Bigagli2024}, a plethora of exciting new platforms are emerging. In this context, the manipulation of magnetic dipolar species such as erbium (Er) and dysprosium (Dy) has been particularly stimulating. This includes the observation of the roton instability \cite{Chomaz2018}, the control of infinite-range interactions in the synthetic dimension of dysprosium atoms \cite{Makhalov2019}, and the observation of supersolidity \cite{ Guo2019, Chomaz2019, Tanzi2019, BoettcherPRX19, Norcia2021, Schmidt2021, Recati2023, Biagioni2024}.

The possibility to create mixtures of dipolar gases opens a new route for studying the interplay between magnetic interactions, quantum fluctuations, and thermal effects. In these systems, a new interaction length scale arises due to interspecies interactions, similar to what is observed in alkali species that has led, for instance, to the observation of quantum liquid droplets \cite{Petrov2015, Cabrera2018}. Such mixtures are poised to exhibit a rich, as-yet-unexplored phase diagram \cite{Goral2002}, with the emergence of self-bound droplets \cite{Arazo2023} and spin-textured supersolids with multiple density-dependent spatial configurations \cite{Goral2002, Saito2009, Xi2018, Arazo2023, Lee2024}. Notably, binary dipolar Bose gases present a supersolid phase whose robustness is independent of quantum fluctuation effects, thus extending the parameter range, in terms of atom number, over which the supersolid clusters are stabilized \cite{Li2022, Bland2022, Scheiermann2023}.

The simultaneous manipulation of ultracold samples of erbium and dysprosium \cite{Trautmann2018} holds promise for probing this rich physics, albeit at the cost of increased experimental complexity. A simpler alternative is the manipulation of dipolar spin mixtures \cite{Santos2006, Burdick2016, Lepoutre2018, Baier2018, Claude2024,Baroni2024}. However, the study of these mixtures has been hampered by two main factors. First, the large magnetic moment usually results in rapid dipolar relaxation processes for all Zeeman sublevels except the lowest energy one, rendering these systems intrinsically unstable \cite{Hensler2003, Pasquiou2010, Burdick2015}. Second, the apparently chaotic nature of the pairwise interactions in erbium and dysprosium, primarily due to their large anisotropic van der Waals coefficients \cite{Kotochigova2014, Frisch2014, Maier2015}, complicates the prediction of interaction sweet spots that would allow the exploration of different miscibility regimes in these mixtures. To date, only the interaction properties of the lowest-energy sublevel, for both erbium and dysprosium, have been characterized \cite{Tang2015, Tang2016, Bottcher2019, Khlebnikov2019, Patscheider2022,Krstajic2023, Lecomte2024}. Efforts have been made to mitigate these factors, either through the preparation in a low magnetic field environment \cite{Lepoutre2019} or strong confinement in deep optical lattices \cite{Barral2024}.

  \begin{figure}[t!]
    \centering
	\includegraphics{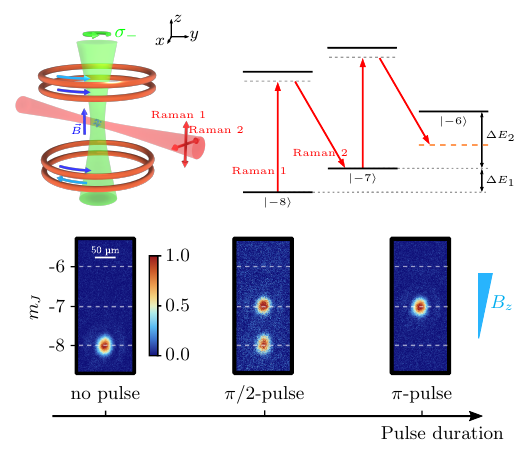}
	\caption{Schematic representation of the experimental protocol. Top panels: two horizontal laser beams (Raman 1 and Raman 2) induce a Raman transition between nearest Zeeman sublevels. 
The vertical laser beam induces a spin-dependent light shift, allowing us to selectively couple the two lowest-energy Zeeman sublevels. Orange line represents the energy of $\ket{-6}$ in the absence of the light-induced quadratic Zeeman effect.
	Bottom panel: absorption images of Bose-Einstein condensates in different internal states, captured after time-of-flight (TOF) expansion in the presence of a magnetic field gradient. The right-most absorption image corresponds to a BEC preparation in $\ket{-7}$ with purity $> 95\%$. Dashed lines serve as guides to the eye for the spatial position of atoms in states $\ket{-8}$, $\ket{-7}$, and $\ket{-6}$.}
    \label{figillustration}
\end{figure}



In this Letter, we introduce a novel binary dipolar quantum gas composed of particles in the Zeeman sublevels $\ket{J=8,\, m_J=-8}$ and $\ket{J=8,\, m_J=-7}$ of $^{162}$Dy (simply labeled hereafter as $\ket{-8}$ and $\ket{-7}$). We identify a magnetic field ``sweet spot'' where dipolar relaxation is suppressed by two orders of magnitude compared to the Wigner threshold law \cite{Weiner1999}. 
Combining our experimental results with a theoretical model of atomic interactions in this regime, and through the analysis of the BEC size scaling with atom number after time-of-flight (TOF) expansion, we determine the intra-species scattering length $a_{77} = 110 (10) \, a_0$ and the interspecies scattering length $a_{78} = 40 (20) \, a_0$, where $a_0$ denotes the Bohr radius. 
 These values, in combination with the already known $a_{88}=\SI{140}{}\, a_0$ \cite{Tang2016, Bottcher2019} allow one to predict the stability diagram for the 7-8 mixture. Additionally, near this optimal magnetic field, we observe spin-dependent Feshbach resonances. These resonances provide precise control over the scattering lengths \cite{ChinRMP10}, enabling the exploration of miscible-immiscible phases in binary dipolar condensates \cite{Wilson2012, Kumar2017, Halder2023, Scheiermann2023}.



We first use an ultracold, but non condensed, sample of $^{162}$Dy. The temperature $T= \SI{250}{\nano\kelvin}$ is chosen to be high enough to avoid possible demixing effects. The gas is confined in an infrared, far-detuned, crossed dipole trap with angular frequencies $\px{\omega_x,\, \omega_y ,\, \omega_z} = 2\pi \times \px{38,\, 212,\, 172} \, \text{Hz}$. The samples are initially prepared in the lowest Zeeman sublevel $\ket{-8}$, in the presence of a magnetic field bias $\mathbf{B} = B \hat{z}$, and contain $\sim 10^5$ atoms (see Ref.~\cite{Lecomte2024} for details).
Taking advantage of the non-zero tensorial part of the polarizability, we create a spin-dependent light shift via a laser beam (see Fig.~\ref{figillustration}), with radius at $1/e^2$ of \SI{100}{\micro\meter}, propagating along the $\hat{z}$ direction, with $\sigma^-$-polarization, blue detuned by $\Delta = 2\pi \times \SI{2.5}{\giga\hertz}$ from the $J'=J-1$ optical transition at $\lambda = 530.305$ nm \cite{Wickliffe2000, li2017} (see also \cite{SuppMat}).
As a result, in the ground-state manifold, the energy of the different Zeeman sublevels is given up to a constant by
\begin{equation}
E(m_J)= \alpha m_J + \gamma (m_J +7)(m_J +8)  \, ,
\end{equation}
where $\alpha = g_J \mu_{\rm B} B$ is the Zeeman energy shift, $g_J$ the Land\'e factor and $\mu_{\rm B}$ the Bohr magneton. We typically work with an optical power of \SI{200}{\milli\watt} such that $\gamma / h \approx 30$ kHz. 
The chosen polarization ensures that the two lowest-energy Zeeman states $\ket{-7}$ and $\ket{-8}$ are uncoupled to the excited manifold, defining them as ``dark states''.

This non-linear energy shift enables precise manipulation of the atomic sample between the $\ket{-8}$ and $\ket{-7}$ states. For this purpose, we use a two-photon Raman process facilitated by two co-propagating laser beams with linear and orthogonal polarizations, detuned from the \SI{626.1}{\nano\meter} atomic transition \cite{Ravensbergen2018b, Chalopin2018}, with Rabi frequency $\Omega_{\rm R} \approx 2\pi\times \SI{23}{\kilo\hertz}$. By performing a Rabi oscillation with adjustable duration, we can prepare either quasi-pure samples in $\ket{-7}$ or spin mixtures with adjustable amplitudes \cite{Burdick2016,Claude2024}. In Fig.~\ref{figillustration}, we present examples of absorption images of Bose-Einstein condensates (BECs) in different internal states, captured after time-of-flight in the presence of a magnetic field gradient.



\begin{figure}
	\center
	\includegraphics{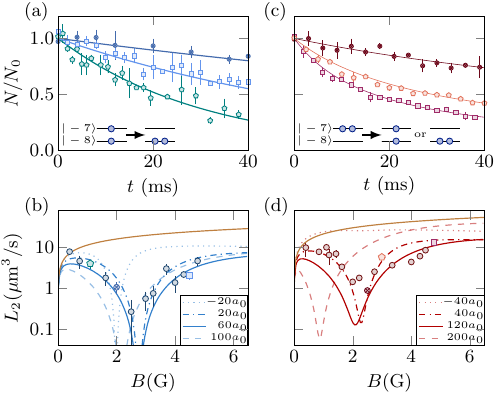}
	\caption{Dipolar relaxation. Time evolution of the atom number in state $\ket{-7}$ for (a) minority component in $\ket{-7}$ ($<5\%$) immersed in the majority component $\ket{-8}$ (blue pentagon $B = \SI{1.1}{\gauss}$, blue crossed circle $B = \SI{2.0}{\gauss}$, and blue square $B=\SI{4.5}{\gauss}$) and for (c) pure sample in $\ket{-7}$ (red crossed circle $B = \SI{2.5}{\gauss}$, red pentagon $B = \SI{3}{\gauss}$ and red square $B=\SI{4.77}{\gauss}$). 
	(b) Two-body loss rate as a function of $B$ for the case of a minority component in $\ket{-7}$. The lines correspond to theoretical predictions (see main text) for scattering length: $a_{ 78} = -20\, a_0$  (blue dotted line), $a_{ 78} = 20\, a_0$  (blue dash dot line), $a_{ 78} = 60\, a_0$  (blue line) and $a_{ 78} = 100\, a_0$  (dashed blue line).
	(d) Two-body loss rate as a function of $B$ for the case of a pure sample in $\ket{-7}$. The lines correspond to theoretical predictions for scattering length: $a_{77} = -40\, a_0$  (red dotted line), $a_{77} = 40\, a_0$  (red dash dot line), $a_{77} = 120\, a_0$  (red line) and $a_{77} = 200\, a_0$  (red dashed line). The brown lines correspond to the Wigner law $\propto \sqrt{B}$.
	}
	\label{figdipolarrelaxation}
\end{figure}

Dipolar relaxation is the primary limitation in manipulating spin mixtures of strongly dipolar atomic gases. It is particularly pronounced in dysprosium due to its large magnetic moment \cite{Chomaz2023}. At non-zero magnetic fields, only the state $\ket{-8}$, which is the lowest-energy Zeeman sublevel, is protected against dipolar relaxation; particles occupying any other internal state will eventually relax towards $\ket{-8}$. Let us consider the simplest relaxation process, where two particles collide, one in the internal state $\ket{-8}$ and the other in $\ket{-7}$. After collision, the particle initially in $\ket{-7}$ can spin-flip towards $\ket{-8}$, a process which we schematically represent by (see Fig.~\ref{figdipolarrelaxation}a)
\begin{equation}
\label{eqrelaxation}
\ket{-7} + \ket{-8} \rightarrow \ket{-8} + \ket{-8}  \, .
\end{equation}
Due to momentum conservation, the two particles equally share the released Zeeman energy in the center-of-mass (CoM) frame. 
For the magnetic fields used in this work, the released energy is much larger than the trap depth, resulting in the loss of both particles; for instance at $B=\SI{1}{\gauss}$, $\Delta E \approx \kb \times \SI{80}{\micro\kelvin}$, where $\kb$ is the Boltzmann constant.

Experimentally, we probe this relaxation process by transferring a small fraction ($< 5\%$) of the ultracold sample into the internal state $\ket{-7}$ and holding the sample for a specified time. Subsequently, we measure the populations of both internal states $\ket{-7}$ and $\ket{-8}$, extracting the two-body loss rate $L_2^{\text{mix.}}$ from the time evolution of the minority component (see Fig.~\ref{figdipolarrelaxation}a). We ensure that over the probed time, both the population and temperature of the atomic sample in $\ket{-8}$ remain constant within $\SI{20}{\%}$.
As shown in Fig.\ref{figdipolarrelaxation}b, we observe a non-monotonic evolution of $L_2^\text{mix.}$ with $B$ over the range of 0.4 to \SI{6}{\gauss}. Remarkably, at $B \approx \SI{2.5}{\gauss}$, the measured loss rate is two orders of magnitude lower than the Wigner law prediction, which scales with magnetic field as $\sqrt{B}$ \cite{Weiner1999,Hensler2003, Burdick2015} (brown straight line in Fig.\ref{figdipolarrelaxation}b).

To explain this spectacular reduction, we first recall the theoretical description, based on the Fermi Golden  Rule (FGR) approach, of a dipolar relaxation process occurring between two atoms $a$ and $b$. This process is induced by the dipolar interaction potential $V_{\rm dd}$, which is proportional to the scalar product of two rank-two tensor operators, ${\cal J}^{(2)}$ and ${\cal U}^{(2)}$, acting on the spin and orbital degrees of freedom of the two-particle system, respectively \cite{Cohen1986,Edmonds1996}:
\begin{equation}
V_{\rm dd}=-\frac{\alpha}{r^3} \sum_{m=-2}^2 (-1)^m {\cal J}^{(2)}_{-m}   \,{\cal U}^{(2)}_{m} \, .
\label{eq:Udd}
\end{equation}
The rank-two tensor operator ${\cal J}^{(2)}=(\bs J_a\otimes\bs J_b)^{(2)}$ is  formed from the two rank-one spin operators $\bs J_a$ and $\bs J_b$. The rank-two orbital operator is defined by ${\cal U}^{(2)}=(\bs u\otimes\bs u)^{(2)}$, with $\bs u=\bs r/r$  where $\bs r=\bs r_a-\bs r_b$ is the relative position variable. We set $\alpha=(3/8)^2 \hbar^2 a_{\rm dd}/M$, where $M$ is the atomic mass and $a_{\rm dd}=129.2\,a_0$ is the so-called dipolar length \cite{Chomaz2023},  characterizing the strength of the dipolar interaction.

A single spin-flip process, which is the only one energetically allowed for the collision of Eq.~\eqref{eqrelaxation}, corresponds to the term $m=1$ in the sum of Eq.~(\ref{eq:Udd}). This term couples the initial s-wave scattering state $\psi_{ i}(\bs r)$ to a final  d-wave state $\psi_{ f}(\bs r)$. The corresponding rate calculated using FGR reads \cite{Moerdijk1996}
\begin{equation}
L_{2}^{\text{mix.}} =\beta k_f \pr{\int \text{d}r \frac{1}{r} \chi_i(r) \chi_f(r)}^2 \, ,
\label{eqL2}
\end{equation}
where $\beta= 54 \pi  \hbar a_\text{dd}^2/(5M)$, $\chi_{i/f}$ stand for the (real) radial parts of $\psi_{i/f}$ and  $k_f =\sqrt{M \Delta E}/\hbar \propto \sqrt{B}$, where $\Delta E$ is the energy released in the spin flip process. In Eq.~(\ref{eqL2}) the wave functions $\chi_{i/f}$ are normalized such that in the absence of interactions between $a$ and $b$ in input and output channels and in the limit of a zero initial energy, we have $\chi_i = 1$ and $\chi_f (r) = j_2(k_f r)$, where $j_2$ is the second spherical Bessel function of the first kind. We note that in this case, the integral entering in Eq.~(\ref{eqL2}) does not depend on $k_f$, which leads to the Wigner threshold law $L_2^{\text{mix.}} \propto \sqrt{B}$.

Interactions in the input and output channels thus play a key role to understand the spectacular reduction of the rate $L_{2}^{\text{mix.}}$ observed in the experiment. Because of van der Waals and dipolar interactions, the wave functions $\chi_{i/f}(r)$ have several nodes in the region where $V_{\rm dd}$ is significant. A variation of the magnetic field $B$ results in a shift of the nodes of $\chi_f$ with respect to those of $\chi_i$ and the integral entering Eq.~\eqref{eqL2} may thus vanish  for a specific value of $B$ (see also \cite{SuppMat}). Physically, this integral can be viewed as the sum of the amplitudes of the paths $\chi_i\to \chi_f$, each path being labelled by the position $r$ at which the spin flip process takes place. The situation for which the integral vanishes corresponds to a globally destructive interference between all these paths.

The full determination of the initial and final wave functions $\psi_{i/f}$ is a complex problem since the dipole interaction mixes all partial wave channels, leading to an infinite set of coupled-channel Schr\"odinger equations \cite{Yi2001,Bohn2009, Kotochigova2014}. Here we adopt a semi-quantitative modeling with only one channel for the initial ($\ell=0$) and for the final ($\ell=2$) states. For each channel, we write the interaction potential $V(r)=V_{\rm cent}(r)-C_6/r^6 +\bar V_{\rm dd}(r)$.
The first term is the centrifugal energy $\hbar^2\ell(\ell+1)/Mr^2$, the second one corresponds to the van der Waals interaction with $C_6 = 2003 \, E_h a_0^6$ \cite{Maier2015}, with $E_h$ being the Hartree energy, and the third term represents the leading effect of dipolar interaction at long distance. For the output channel, we use $\bar V_{\rm dd}(r)=-C_3/r^3$ with $C_3=(6/7) \hbar^2 a_{\rm dd}/M$, which represents the angular average of $V_{\rm dd}(\bs r)$ for the considered partial wave. For the input channel, this angular average vanishes. We therefore consider the next-order term, $\bar V_{\rm dd}(r)=-C_4/r^4$ with $C_4=(147/160)\hbar^2 a_{\rm dd}^2/M$. This term results from second-order perturbation theory, considering the coupling between the input s-wave state and the d-wave states of the same spin multiplicity, with their splitting determined by the centrifugal energy [for details see \cite{Maier2015b} and references therein].  Note that the dipolar length  $a_{\rm dd}$ is comparable to the length scale associated with the van der Waals interaction $R_6=(MC_6/\hbar^2)^{1/4}=156\,a_0$ \cite{Gao2008}. Our perturbative treatment of $V_{\rm dd}$  holds for $r>a_{\rm dd}\approx R_6$. For smaller values of $r$, dipolar interactions should be handled in a non-perturbative manner, but they are small compared to van der Waals interactions and are expected to play a lesser role.

We numerically compute $\chi_\text{i}$ and $\chi_\text{f}$ by imposing a hard-core potential at short distances in the resolution of the Schr\"{o}dinger equation, such that the total number of bound states equals the value $71$ predicted in Ref.~\cite{Petrov2012}. Quasi-identical results are obtained for $L_2^{\text{mix.}}$ with a Lennard-Jones potential with the same number of bound states, and also when the  number of bound states is varied by $\pm 10\%$.

The evolution of  $L_2^{\text{mix.}}$ with the magnetic field strongly depends on the interstate scattering length $a_{78}$. As shown in Fig.\ref{figdipolarrelaxation}b, we find good agreement with our experimental data for $a_{78} = 20$--$60 \, a_0$.
Furthermore, our results effectively exclude negative scattering length values or large positive values, $a_{78} \gtrsim 100 \, a_0$, providing crucial information for determining the stability and miscibility regime of these samples. Note that our result does not verify $a_{78}  = a_{88}$, as one would expect for pure contact interactions \cite{Kawaguchi2012}.

We apply the same method to the case of  pure $\ket{-7}$ samples. In this case, we fit the atom number evolution by $N_0/\pc{1 + L_2 \bar{n} t}$, where $\bar{n}$ is the initial sample-averaged density and $N_0$ the initial atom number (see Fig.~\ref{figdipolarrelaxation}c). As shown in Fig.~\ref{figdipolarrelaxation}d, we also observe a non-monotonic evolution of $L_2^\text{pol.}$ with $B$. In this case, there are two possible final channels, $\ket{-7} + \ket{-8}$ and $\ket{-8} + \ket{-8}$, each leading to a contribution analogous to Eq.(\ref{eqL2}). Therefore we do not expect that there exists a $B$ field for which the total rate $L_2$ exactly cancels, which makes the determination of $a_{77}$ by this method less accurate than for $a_{78}$. Using the same methodology, generalized to the case of losses resulting from the collision between two particles in state $\ket{-7}$ (see \cite{SuppMat}), we extract the possible range for the scattering length $a_{7 7} = 40$--$120 \, a_0$.

Interestingly, both loss rate minima are located at approximately the same magnetic field, $B=\SI{2.5}{\gauss}$, and equal to $L_{2,\, \text{min}}^\text{mix.} = \SI{0.26 (25)}{\micro\meter}^3/\SI{}{\second}$ and $L_{2,\, \text{min}}^\text{pol.} = \SI{0.9 (2)}{\micro\meter}^3/ \SI{}{\second}$. These values correspond to an almost two-order of magnitude reduction compared to the Wigner law. \cRL{Furthermore, the minimal value $L_{2,\, \text{min}}^\text{mix.}$ is likely an overestimate, as it is derived from long probing times, during which one-body losses may contribute to the overall observed losses.} At this magnetic field, we also measure the two-body relaxation rate for a pure BEC in $\ket{-7}$ and find $L_{2,\, \text{BEC}} = \SI{0.3(1)}{\micro\meter}^3/\SI{}{\second}$, consistent with the expected twofold reduction due to the decreased two-body correlation function, at short range, for a BEC.



  \begin{figure}[t!]
	\center
	\includegraphics{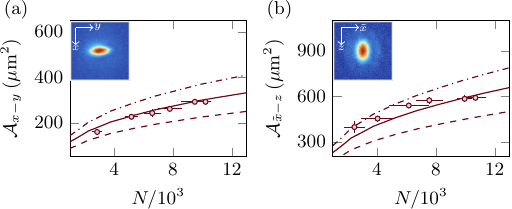}
	\caption{Determination of the scattering length from time-of-flight expansion of a pure BEC in $\ket{-7}$ at a magnetic field $B = \SI{1.43}{\gauss}$. 
	Area of the BEC, defined as the product of the Thomas-Fermi radii extracted from inverted parabola fits to absorption images, plotted as a function of atom number $N$ in the (a) $x$--$y$ and (b) $\tilde{x}$--$z$ planes, with $\tilde{x}$ at a $70^\circ$ angle from ${x}$ and perpendicular to $z$. The lines correspond to numerical simulations for the BEC expansion, assuming scattering lengths of $a_{77} = 110 \, a_0$ (solid line), $a_{77} = 98 \, a_0$ (dashed line), and $a_{77} = 122 \, a_0$ (dot-dash line). \cRL{The trapping frequencies are $\px{\omega_x,\,\omega_y,\,\omega_z} = 2\pi \times \px{100,\, 260,\, 200} \text{ Hz}$.}}
	\label{figscatt}
\end{figure}

To confirm and improve on our determination of the scattering length $a_{77}$, we prepare a quasi-pure BEC in the internal state $|-7\rangle$ and analyze its expansion when it is released from the trap. The release occurs immediately after the transfer from $|-8\rangle$ to $|-7\rangle$ and we record the atomic density  after a 13.1\,ms TOF. Absorption imaging along two orthogonal axes allows us to observe the cloud expansion in all spatial directions. We show in Fig.~\ref{figscatt} the variation of the cloud sizes after TOF as function of the atom number. The continuous lines show the prediction of a model based on the Castin-Dum scaling equations \cite{Castin1996}, here generalized to take into account dipolar interactions \cite{Odell2004, Eberlein2005} (see also \cite{SuppMat}). The comparison between numerical and experimental results gives $a_{77} = 110 (10) \, a_0$, which is in agreement with the value inferred from the two-body loss rate (Fig.\ref{figdipolarrelaxation}). As a sanity check, we repeated this experiment for a BEC prepared in $|-8\rangle$ and obtained $a_{88} = 136(8) a_0$, which agrees with published values \cite{Tang2016, Bottcher2019}. Similarly we have measured the scattering length $a_{66} = 92 (10) \,a_0$ (see \cite{SuppMat}). 

The determination of $a_{77}$, $a_{88}$, and $a_{78}$ allows us to draw conclusions about the miscibility regime of this spin mixture. In the absence of dipolar interactions, the scattering length values would indicate that a mixture with equal populations in $\ket{-8}$ and $\ket{-7}$ is miscible, as it satisfies the inequality $a_{88} \, a_{77} \geq a_{78}^2$. However, the anisotropy of dipole-dipole interactions makes this conclusion more subtle. 
For a homogeneous 3D gas, we predict from the measured values of the three scattering lengths  that the mixture is non miscible, because the head-to-tail arrangement of dipoles favors spatial separation of the two spin components \cite{Wilson2012}. 
\cccRL{In contrast, applying a strong confinement along the dipole orientation, should lead to miscible mixtures \cite{Kumar2017}. Taking into account the reported scattering length values and our trap frequencies, it appears our system already lies in the miscible regime \cite{Scheiermann2024}}.



\begin{figure}[t!]
    \centering
    \includegraphics{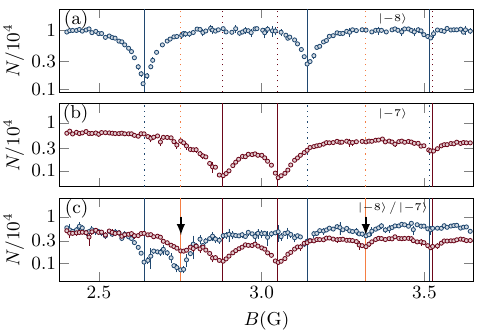}
    \caption{Spin-dependent loss features following a \SI{40}{\milli\second} hold time at the target magnetic field. (a) Population variation with magnetic field for a pure BEC in $\ket{-8}$. (b) Population variation for a pure BEC in $\ket{-7}$. (c) Population variation for the case of a 50-50 spin mixture in $\ket{-8}$ (blue) and $\ket{-7}$ (red). The vertical lines represent the different spin-dependent Feshbach resonances. The vertical black arrows point to the interspecies loss features. }
    \label{figFFRm7}
\end{figure}

Alternatively, changing one of the aforementioned scattering lengths also gives access to different miscible regimes. To explore this aspect, we have identified spin-dependent Feshbach resonances near the dipolar relaxation minimum at $B \approx \SI{2.5}{\gauss}$. These resonances occur in a magnetic field region where dipolar relaxation processes are strongly suppressed (see \cite{SuppMat}), ensuring the validity of our approach for exploring new phases of matter.
We show in Fig.\ref{figFFRm7} the fraction of atoms remaining in the trap after a hold time of \SI{40}{\milli\second}, taking as initial state a pure BEC with adjustable fractions of $\ket{-7}$ and $\ket{-8}$. When all atoms are in $\ket{-8}$ (Fig.\ref{figFFRm7}a), we recover the known three loss features described in Ref.\cite{Lecomte2024} for this magnetic field range. When all atoms are in $\ket{-7}$ (Fig.\ref{figFFRm7}b), we observe again three loss resonances, located at magnetic fields that differ from those for $\ket{-8}$. 

The case of a $50\%-50\%$ spin-mixture case reveals extra features (Fig.\ref{figFFRm7}c). In addition to the resonances observed for pure \ket{-8} and \ket{-7} BECs, we observe two new features affecting both spin components. We interpret these losses as three-body recombination processes resulting from interspecies Feshbach resonances, involving two particles in a given spin state and another particle in a different spin state. We thus expect that the losses for the two states in an initial 50-50 mixture are at most in a ratio 2:1, and reach equal values if the two processes 7+7+8 and 7+8+8 have equal probabilities. Here we find a ratio 1.5 and 0.8 between the losses for state \ket{-8} and state \ket{-7} for the two resonances of Fig.\ref{figFFRm7}c at 2.75 G and 3.3 Gauss, respectively, which is compatible with the expected bound. Interestingly, these two resonances are located at magnetic fields equidistant from nearby intraspecies Feshbach resonances for $\ket{-7}$ and $\ket{-8}$.



In conclusion, our study represents a significant advancement in the preparation and stabilization of binary dipolar gases. We have identified a magnetic field region where two-body losses due to dipolar relaxation are reduced by nearly two orders of magnitude compared to the Wigner threshold law, making this system long-lived, similarly to its fermionic counterpart \cite{Burdick2015, Burdick2016}. 
This result has allowed us to estimate both intra- and inter-species scattering lengths using a model based on single-channel scattering theory. A natural extension of our work is to develop a multi-channel approach \cite{Kotochigova2014, Secker2021} to improve the accuracy of the determination of these scattering lengths from the measured relaxation rates.
The exceptionally low two-body loss rate, observed near 2.5 G in a spin mixture of $\ket{-7}$ and $\ket{-8}$, combined with the identification of multiple inter- and intra-species Feshbach resonances near this magnetic field, opens the possibility of studying various miscible regimes within this binary dipolar mixture. This development paves the way for exploring new quantum phases in dipolar gases.

\begin{acknowledgements}
\textit{Acknowledgements: } We acknowledge fruitful discussions with the members of the Bose-Einstein condensate team at LKB, Bruno Laburthe-Tolra and Luis Santos. We thank Sylvain Nascimbene for a careful reading of the manuscript, and Francesca Ferlaino for stimulating discussions and for identifying an issue with the labeling in the previous version of Fig.~2.
This research was funded, in part, by l'Agence Nationale de la Recherche (ANR), projects ANR-20-CE30-0024 and ANR-24-CE30-7961. For the purpose of open
access, the author has applied a CC-BY public copyright licence to any Author Accepted Manuscript (AAM) version arising from this submission.
This work was also funded by Region Ile-de-France in the framework of DIM QuanTiP.
\end{acknowledgements}


\bibliographystyle{ieeetr}
\bibliography{bib.bib}


\cleardoublepage

\renewcommand{\thefigure}{S\arabic{figure}}
\setcounter{figure}{0} 
\renewcommand{\theequation}{S\arabic{equation}}
\setcounter{equation}{0} 

\appendix
\onecolumngrid 

\section*{Supplemental Material}

%
\twocolumngrid
\section{Experimental details}
\label{secRaman}

\subsection{Characterization of the optical transition at \SI{530.3}{\nano\meter} }
\label{greenlaser}

\begin{figure}[t!]
	\center
	\includegraphics{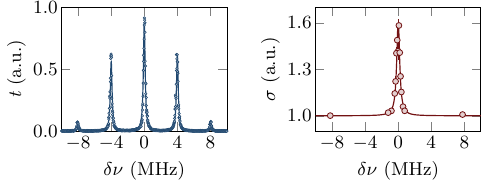}
	\caption{Characterization of the green transition. {Left panel:} transmission response ($t$) through a ULE cavity with a finesse of 300 000. Sidebands at \SI{4}{\mega\hertz} are added to characterize the laser linewidth. {Right panel:} size of an atomic sample after a \SI{6}{\milli\second} time-of-flight expansion following exposure to a near-resonant laser beam for \SI{40}{\micro\second}.}
	\label{figGreen}
\end{figure}

The spin-dependent light shift is created via a laser beam propagating along the vertical direction with $\sigma^-$ polarization. The laser is blue-detuned from the $J' = J-1$ transition corresponding to the excited state 4f$^{10}(^6$F$)$5d6s$^2$.
We characterize this transition by performing a ``push'' experiment \cite{Chalopin2018}, which involves applying a $\SI{40}{\micro\second}$ long near-resonant beam that is circularly polarized ($\sigma^+$) and propagates along the $\hat{z}$ axis with a saturation parameter $s = I/I_\text{sat} \approx 0.5$, where $I_\text{sat}$ is the saturation intensity. In the limit of a short duration pulse, the momentum kick experienced by the atoms reaches its maximum value when the laser frequency matches the transition frequency. This results in a broadening of the momentum distribution of the atomic sample, $\sigma$, after time of flight.
In Fig.\ref{figGreen} (right panel), we show the deformation of the ultracold sample as a function of the laser frequency of the near-resonant laser beam. We deconvolve the response function by the linewidth of the laser beam, measured using a high-quality factor ultra-low expansion (ULE) cavity (see Fig.\ref{figGreen}, left panel). From this, we deduce that the natural linewidth of the optical transition is \SI{170(20)}{\kilo\hertz}, which is in good agreement with the NIST database \cite{Wickliffe2000}.

\subsection{Inducing the Raman transitions}

The spin-dependent light shift allows us to selectively couple the two lowest-energy states, $\ket{-8}$ and $\ket{-7}$. For that purpose, we use a two-photon Raman process, achieved by two co-propagating laser beams with linear and orthogonal polarizations, oriented at a $70^\circ$ angle relative to the $\hat{x}$-axis in the $x-y$ horizontal plane. The laser frequency is red-detuned by $\Delta / 2\pi = \SI{-90}{\giga\hertz}$ with respect to the atomic transition at \SI{626.1}{\nano\meter}, associated with the excited level 4f$^{10}(^5$I$_8)$6s6p($^3$P$_1^\circ$) (8,1)$_9^\circ$ with quantum number $J'=J+1$ \cite{Ravensbergen2018b}.

The two-photon Rabi frequency is chosen to be approximately $\Omega_{\rm R} \approx 2\pi \times \SI{23}{\kilo\hertz}$, hence a $\pi$-pulse duration of $ \pi/ \Omega \approx \SI{22}{\micro\second}$. This method allows the preparation of pure samples in $\ket{-7}$ with fidelity greater than $95\%$, as well as spin mixtures with varying amplitudes.

The preparation of a pure Bose-Einstein Condensate (BEC) in $\ket{-6}$ is performed similarly, but using a laser beam detuned by \SI{-100}{\giga\hertz} from the \SI{626.1}{\nano\meter} transition to create the spin-dependent light shift, instead of the \SI{530.305}{\nano\meter} optical transition, since $\ket{-6}$ is coupled to the excited manifold with quantum number $J'=J-1$, and therefore does not correspond to a ``dark state''. The atomic sample is first transferred from $\ket{-8}$ to $\ket{-7}$ with a $\pi$-pulse and then from $\ket{-7}$ to $\ket{-6}$ with a second $\pi$-pulse. The laser frequency of one of the two beams is adjusted between the two pulses to ensure the resonance condition for each process. The transfer into $\ket{-6}$ lasts for \SI{50}{\micro\second}, and the sample has a purity $\gtrsim 92\%$.

\section{Two-body loss rate}
\label{sec2bodylosses}

We provide a more detailed description of the two-body loss mechanism related to dipolar relaxation. This discussion is divided into two subparts: the relaxation of an impurity in $\ket{-7}$ in a bath of $\ket{-8}$, and collisions between particles in $\ket{-7}$. In our case, using a spin-dependent light shift, the energy difference between the internal states $\ket{-8}$ and $\ket{-7}$ differs from the energy difference between $\ket{-7}$ and $\ket{-6}$. Therefore, spin-exchange collisions $\ket{-7 } + \ket{-7} \rightarrow \ket{-6} + \ket{-8}$ are suppressed by energy conservation \cite{Claude2024}.

\subsection{Expression of the dipole-dipole coupling}

The magnetic dipole-dipole interaction between two particles  with magnetic moments $\bs \mu_a$ and $\bs \mu_b$ and coordinates $\bs{r}_a$ and $\bs{r}_b$ is
\begin{equation}
V_{\rm dd} = \frac{\mu_0}{4\pi r^3}\pr{\bs \mu_a \cdot \bs \mu_b - 3\pc{\bs \mu_a \cdot \bs u} \pc{\bs \mu_b \cdot \bs u}}
\end{equation}
with $\bs r=\bs r_a-\bs r_b$ and $\bs u=\bs r/r$. We use the link between the magnetic moment $\bs \mu_i$ ($i=a,b$) and the spin $\bs J_i$ as $\bs \mu_i=g_J\mu_{\rm B}\bs J_i$, where $ g_J$ the Land\'e factor and $\mu_{\rm B}$ the Bohr magneton, and we define the dipole length
\begin{equation}
a_{\rm dd}= \frac{M\mu_0 (g_J\mu_{B}J)^2}{12 \pi \hbar^2}.
\end{equation} 
The dipole-dipole interaction can then be written as \cite{Edmonds1996}
\begin{equation}
V_{\rm dd} =-\frac{9}{J^2}\,\frac{\hbar^2 a_{\rm dd}}{Mr^3 }
\pr{ \pc{\bs J_a \cdot \bs u} \pc{\bs J_b\cdot \bs u}-\frac{1}{3} \bs J_a\cdot \bs J_b}.
\label{eq:Vdd_9}
\end{equation}
The term in bracket can be written as 
\begin{equation}
\pc{\bs J_a \cdot \bs u} \pc{\bs J_b \cdot \bs u}-\frac{1}{3} \bs J_a\cdot \bs J_b= \sum_{m=-2}^2 (-1)^m {\cal J}^{(2)}_{-m}   \,{\cal U}^{(2)}_{m} 
\end{equation}
where we have introduced spin ${\cal J}^{(2)}$ and orbital ${\cal U}^{(2)}$ rank-two tensor operators, so that Eq.~\eqref{eq:Vdd_9} coincides with Eq.~\eqref{eq:Udd} for $J=8$.  

The components of the spin operator ${\cal J}^{(2)}$ are defined as
\begin{eqnarray}
{\cal J}^{(2)}_{\pm 2}&=&J_{a,\pm 1}J_{b,\pm 1} \\
{\cal J}^{(2)}_{\pm 1}&=&\frac{1}{\sqrt 2}\left(J_{a,\pm 1}J_{b,0} + J_{a,0}J_{b,\pm 1} \right)\\
{\cal J}^{(2)}_{0}&=&\frac{1}{\sqrt 6}\left(J_{a,+1}J_{b,-1}+2J_{a,0}J_{b,0}+J_{a,-1}J_{b,+1}
\right)
\end{eqnarray}
where we have used the standard components of a vector operator $\bs J$ as a function of its cartesian components $J_{x},J_{y},J_{z}$:
\begin{equation}
J_{\pm 1}=\mp\frac{1}{\sqrt 2}\left(J_{x}\pm \mi J_{y}\right) \qquad J_{0}=J_{z}.
\end{equation}
Note that $J_{\pm 1}$ differ from the ladder operators $J_{\pm}=J_{x}\pm \mi J_{y}$ by a factor $1/\sqrt 2$ (and a sign for $J_{+1}$).

The components of the orbital operator ${\cal U}^{(2)}$ are constructed in a similar way starting from the vector operator $\bs u$. They are functions of the spherical coordinates $\theta,\varphi$ of the unit vector $\bs u$ 
\begin{eqnarray}
{\cal U}^{(2)}_{\pm 2}&=&\frac{1}{2}\sin^2 \theta\;e^{\pm 2\mi\varphi}
\label{}\\
{\cal U}^{(2)}_{\pm 1}&=&\mp \sin \theta \cos\theta \,e^{\pm \mi\varphi} \label{} \\
{\cal U}^{(2)}_{0}&=&\frac{1}{\sqrt 6}\left(3\cos^2 \theta -1\right)
\end{eqnarray}
 These components are proportional to the rank-2 spherical harmonics: ${\cal U}^{(2)}_{m}=\sqrt{{8\pi}/{15}}\,Y_{2,m}$.

\subsection{Collisions between $\ket{-7}$ and $\ket{-8}$}

We now consider the dipolar relaxation following the collision between two particles in the internal states $\ket{-7}$ and $\ket{-8}$. As discussed in the main text, the dipolar relaxation rate is given by the Fermi golden rule, which in this case is expressed as \cite{Moerdijk1996}:
\begin{equation}
\label{eqoverlap}
L_2^\text{mix.} = \frac{54 \pi}{5} \frac{\hbar k_f}{M} a_\text{dd}^2 \times \pr{\int \text{d}r\, \frac{1}{r} \chi_\text{i} (r) \chi_\text{f} (r)}^2 ,
\end{equation}
with $k_f = \sqrt{M g_J \mu_{\rm B} B} /\hbar$, $M$ is the atomic mass and $\chi_{\text{i}/\text{f}}$ are the (real) radial parts of the incoming and outgoing collision wavefunctions. 

\begin{figure}
	\center
	\includegraphics{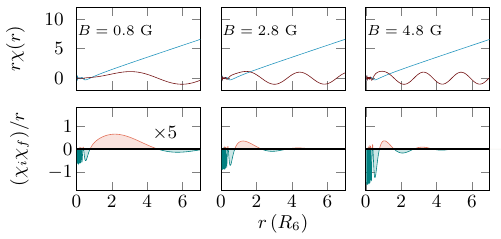}
	\caption{Radial part of the collisional wavefunctions and their overlap. {Top panels:} radial part of the incoming and outgoing collisional wavefunctions for different magnetic fields. The incoming collisional wavefunction has zero energy and the scattering length $a = 60 \, a_0$. {Bottom panels:} spatial dependence of the contribution to the matrix element entering in $L_2$. We show in green and orange the negative and positive contributions to the integral of Eq.~\eqref{eqoverlap}, respectively. The integral cancels for $B \approx 2.8$~G.}
	\label{figWvf}
\end{figure}

To gain an intuitive understanding of the dependence of the integral in Eq.\eqref{eqoverlap} on $B$, we depict the radial parts of the incoming and outgoing wavefunctions, as well as their product divided by $r$, in Fig.\ref{figWvf}. Here, we have chosen a zero-energy incoming wavefunction, leading to an asymptotic behavior $r \chi_i (r) \sim r - a$ with $a = 60 \, a_0$. The outgoing wavefunction is normalized such that its asymptotic behavior is $r \chi_f (r) = \sin(k_f r + \delta)$.
It can be observed that around a magnetic field of $B \approx \SI{2.8}{\gauss}$, the negative and positive parts of the product between the two functions compensate each other, resulting in a complete cancellation of the relaxation rate.

\subsection{Collisions between $\ket{-7}$ and $\ket{-7}$}

In the case of a pure BEC, two decay mechanisms are possible, represented by 
\begin{equation*}
\begin{aligned}
\ket{-7} + \ket{-7} &\longrightarrow \ket{-7} + \ket{-8} \hspace{1cm} \text{(process 1)}\, , \\
\ket{-7} + \ket{-7} &\longrightarrow \ket{-8} + \ket{-8} \hspace{1cm} \text{(process 2)} \, .
\end{aligned}
\end{equation*}

As explained in the main text, for each channel, we write the interaction potential as
\begin{equation}
V(r)=V_{\rm cent}(r)-C_6/r^6 +\bar V_{\rm dd}(r)\, .
\end{equation}
Here, the leading effect of the dipolar interaction at long distances is given by $\bar V_{\rm dd}(r)=-C_4/r^4$ with $C_4=(6/5)\times (7/8)^4 \hbar^2 a_{\rm dd}^2/M$ for the incoming channel, and  $\bar V_{\rm dd}(r)=-C_3/r^3$ for the two possible outgoing channels with
\begin{equation*}
\begin{aligned}
C_3 = \frac{3}{4} \frac{\hbar^2 a_{\rm dd}}{M} \hspace{1cm} &\text{(process 1)}\\
C_3 = -\frac{12}{7} \frac{\hbar^2 a_{\rm dd}}{M} \hspace{1cm} &\text{(process 2)} \, .
\end{aligned}
\end{equation*}

The relaxation rates associated to the two processes are given by
\begin{equation*}
\begin{aligned}
L_2^{(1)} &= g^{(2)}(0)\frac{54 \pi}{5} \pc{\frac{7}{8}}^2 \frac{\hbar k_f}{M} a_\text{dd}^2 \times \pr{\int \text{d}r\, \frac{1}{r} \chi_\text{i} (r) \chi_\text{f,1} (r)}^2 \\
L_2^{(2)} &= g^{(2)}(0)\frac{54 \sqrt{2} \pi}{5} \pc{\frac{1}{8}} \frac{\hbar k_f}{M} a_\text{dd}^2 \times \pr{\int \text{d}r\, \frac{1}{r} \chi_\text{i} (r) \chi_\text{f,2} (r)}^2 \hspace{-0.2cm},
\end{aligned}
\end{equation*}
where $g^{(2)}(0)$ is the local two-body correlation function.
The total relaxation rate is the sum of both terms. Since  $\chi_{f, 1}$ and $\chi_{f, 2}$ have different periodicities due to the factor of 2 difference in the released energy between the two processes, the two rates cancel at different magnetic fields, thereby preventing the cancellation of the sum.

\section{Expansion of a dipolar BEC and determination of scattering length}
\label{appnumerics}

\begin{figure}[t!]
	\center
	\includegraphics{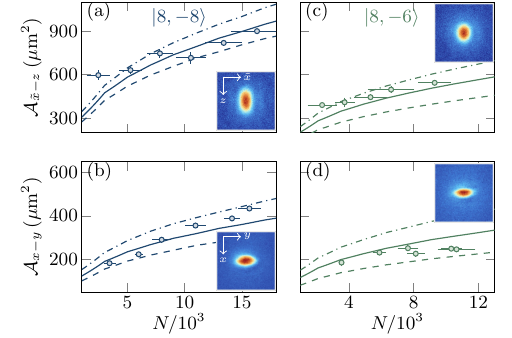}
	\caption{Determination of the scattering length from time-of-flight expansion of a pure BEC in $\ket{-8}$ and $\ket{-6}$ at a magnetic field $B = \SI{1.43}{\gauss}$. 
	{Left:} area of the BEC in $\ket{-7}$, defined as the product of the Thomas-Fermi radii extracted from inverted parabola fits to absorption images, plotted as a function of atom number $N$ in the (a) $x$--$y$ and (b) $\tilde{x}$--$z$ planes. The lines correspond to numerical simulations for the BEC expansion, assuming scattering lengths of $a_{88} = 136 \, a_0$ (solid line), $a_{88} = 128 \, a_0$ (dashed line), and $a_{88} = 144 \, a_0$ (dot-dash line). 
	{Right:} area of the BEC in $\ket{-6}$, plotted as a function of atom number $N$ in the (a) $x$--$y$ and (b) $\tilde{x}$--$z$ planes. The lines correspond to numerical simulations for the BEC expansion, assuming scattering lengths of $a_{66} = 92 \, a_0$ (solid line), $a_{66} = 76 \, a_0$ (dashed line), and $a_{66} = 108 \, a_0$ (dot-dash line). }
	\label{figExpansionm8m6}
\end{figure}

Dipolar interactions contribute in a non-trivial way to the expansion of a dipolar BEC. Determining the scattering length thus requires numerical simulations to explain the evolution of the cloud size as a function of atom number, which we briefly summarize here.
Following the work of Ref.~\cite{Eberlein2005}, we first compute the stationary solution of a dipolar BEC prepared in  $\ket{-8}$, through the minimization of the functional energy 
 \begin{equation}
E [\psi] = \int \text{d}^3r\,  \md{\psi}^2 U
\end{equation}
where
\begin{equation}
U = \frac{1}{2} M \pc{\omega_x^2 x^2 + \omega_y^2 y^2 + \omega_z^2 z^2} + \frac{1}{2} g n(r) +\frac{\Phi_{\rm dd}}{2}\, ,
\end{equation} 
and $n(\vec{r}) = \md{\psi}^2$, with $\omega_i$ the trapping frequency along $\hat{i}$-axis, $g = \frac{4\pi \hbar^2 a}{M}$. We define the spatial density as
\begin{equation}
n(r) = n_0 \pr{1 - \pc{\frac{x}{R_x}}^2 - \pc{\frac{y}{R_y}}^2 - \pc{\frac{z}{R_z}}^2} \, ,
\end{equation} 
with
\cRL{
\begin{equation}
n_0 = \frac{15}{8\pi} \frac{N}{R_x R_y R_z}\, ,
\end{equation}
}
where $R_i$ is the Thomas-Fermi radius along $\hat{i}$. The term $\Phi_{\rm dd}$ amounts for the change in internal energy due to magnetic dipole-dipole interactions. For a generic trap, this term does not display an analytical expression, and we numerically compute it as
\begin{align}
    \Phi_{\rm dd} = -C_{\rm dd} \hat{e}_i \hat{e}_j \pc{\nabla_i \nabla_j \phi (\mathbf{r}) +\frac{\delta_{ij}}{3} n(r)}
\end{align}
where $C_{\rm dd} = (12 \pi \hbar^2 a_{\rm dd})/M$, and 
\begin{align}
\phi(\mathbf r) = \frac{1}{4\pi} \int \text{d}^3 \mathbf r' \frac{n(\mathbf r')}{\md{\mathbf r -\mathbf r'}} \, .
\end{align}
The cloud expansion is then obtained through the resolution of the coupled equations
\begin{align}
    \frac{N m}{7} R_i^2(0) \Ddot{b_i} = -\frac{\partial}{\partial b_i} E(\omega_x = \omega_y = \omega_z =0)
\end{align}
where $R_i (0)$ is the in-trap radius along the direction $\hat{i}$, $b_i(t) = R_i (t)/R_i(0)$ is a dimensional expansion coefficient and $E(\omega_x = \omega_y = \omega_z =0)$ is the functional energy in the absence of a trapping potential. 
For the expansion of a BEC in $\ket{-7}$ and $\ket{-6}$ we factor in the 25\% initial loss due to dipolar relaxation during the expansion. 

We show in Fig.~\ref{figExpansionm8m6} the evolution of the area with atom number $N$ for a BEC prepared in $\ket{-8}$ and $\ket{-6}$. Following the same methodology as for a pure BEC in state $\ket{-7}$ (see main text), the comparison between numerical and experimental results give the scattering lengths $a_{88} = 136 (8) \, a_0$ and $a_{66} = 92 (10) \,a_0.$

\section{Loss features for $\ket{-8}$ and $\ket{-7}$ BECs}

\begin{figure}[t!]
	\center
	\includegraphics{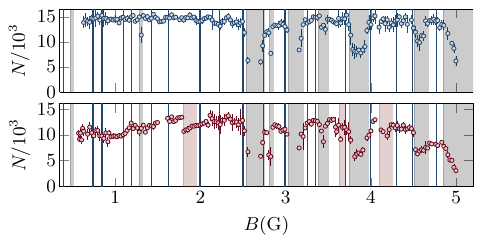}
	\caption{\cRL{Loss features for a pure BEC in $\ket{-8}$ (top panel) and $\ket{-7}$ (bottom panel). The vertical blue lines indicate the magnetic fields at which $L_2$ was measured (see Fig.~\ref{figdipolarrelaxation}). The grey regions highlight magnetic fields where significant losses occur in the $\ket{-8}$ state, while the red  regions mark strong loss features observed for a pure BEC in state $\ket{-7}$.}}
	\label{lossfeatures}
\end{figure}

\cRL{
Before determining $L_2$ at different magnetic fields (see Fig.~\ref{figdipolarrelaxation}), we first identify and exclude magnetic field regions where Feshbach resonances lead to significant three-body losses. These losses could obscure the dipolar relaxation processes that we aim to quantify. As shown in Fig.~\ref{lossfeatures}, we measure the atom number of a pure BEC as a function of magnetic field $B$ over the range of 0.5-5 G. The BEC is initially prepared in the $\ket{-8}$ state at a fixed magnetic field. We then quench the magnetic field to the desired value, hold it for \SI{100}{\milli\second}, and measure the atom number after a \SI{10}{\milli\second} time-of-flight expansion. This procedure reveals several loss features, highlighted as grey regions in the top panel of  Fig.~\ref{lossfeatures}, which are due to three-body recombination of \ket{-8} atoms associated to Feshbach resonances.\\
In a second experiment, we set the magnetic field to the target value and perform a two-photon Raman transfer to probe the loss features of a pure BEC in the \ket{-7} state. After the transfer, we immediately release the cloud to minimize losses due to dipolar relaxation processes. Several new loss features, which are signatures of three-body losses now for \ket{-7}  atoms, emerge. These features are highlighted by the red colored regions in the bottom panel of Fig.~\ref{lossfeatures}.\\
This calibration allows us to carefully select magnetic field values outside the grey and red color regions for probing the dipolar relaxation rate (see vertical blue lines in Fig.~\ref{lossfeatures}), ensuring that our determination $L_2$ (see Fig.~\ref{figdipolarrelaxation}) is not affected by the three-body loss features associated with either the \ket{-8} or \ket{-7} states.
}

\section{Dependence of the Two-Body Loss Rate on the Scattering Length}

\begin{figure}[t!]
	\center
	\includegraphics[width=8.cm]{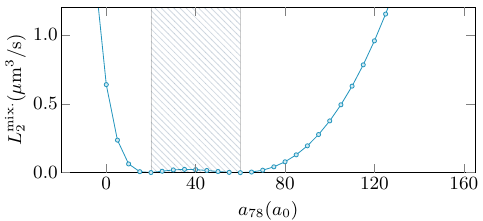}
	\caption{$L^\text{mix.}_2$ as a function of the interspecies scattering length, $a_{78}$, for a magnetic field of \SI{2.75}{\gauss}. The hatched region represents the estimation window for the interpescies background scattering length.}
	\label{figL2scatt}
\end{figure}

As shown in Fig.~\ref{figFFRm7}, we observed both inter- and intra-species Feshbach resonances between 2.5 and 3 G. Since the minimum two-body dipolar relaxation rates depend on the scattering properties of the atomic sample, it is natural to question how the mixture is influenced by changes in the different scattering lengths. To investigate this, we focus on the case of the interspecies Feshbach resonance located at 2.75 G. At this magnetic field, it should be possible to vary the interspecies scattering length $a_{78}$ without modifying the intraspecies scattering lengths $a_{77}$ and $a_{88}$.\\
In Fig. \ref{figL2scatt}, we show the variation of $L^\text{mix.}_2$ as a function of $a_{78}$ at a magnetic field of 2.75 G (blue circles). We find that for scattering lengths ranging from 0 to 125 $a_0$, $L^\text{mix.}_2 \lesssim 1 \,\mu \text{m}^3/\text{s}$, corresponding to a lifetime $\gtrsim $\SI{100}{\milli\second} for a typical BEC density of $n = 10^{19} \mu \text{m}^{-3}$. This lifetime is sufficiently long to study the transition from the miscible to the immiscible regime for a quasi-2D sample, as $a_{78}$ evolves from a negligible value compared to $a_{77}$ and $a_{88}$ to a value exceeding the geometric mean of the intraspecies scattering length background values \cite{Wilson2012}. Furthermore, note that, to a good approximation, $a_{77}$ remains constant over the narrow magnetic field region of the $a_{78}$ resonance, which means that the dipolar relaxation rate involving two particles in $m=-7$ does not change and remains at a low value (see Fig.~\ref{figdipolarrelaxation}).

\section{Determination of $L_2$ from the experimental data}

\begin{figure}[t!]
	\center
	\includegraphics[width=8.8cm]{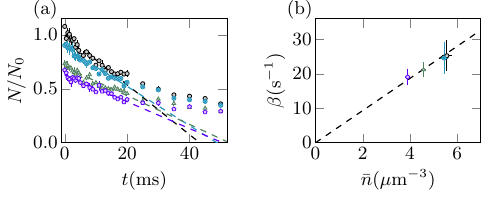}
	\caption{\cRL{Dependence of the loss rate on density. (a) Evolution of the atom number for a pure BEC in the $\ket{-7}$ state, shown for three different initial atom numbers at a magnetic field bias of 4~G. (b) Initial loss rate as a function of the average density. The linear fit demonstrates that the loss rate increases linearly with density, consistent with expectations for two-body loss processes.}}
	\label{figL2Linear}
\end{figure}

We fit the evolution of the atom number using the following function: 
\begin{align} 
N(t) = N_0 \frac{1}{\left(1 + \frac{\alpha-1}{\tau}t\right)^{\frac{1}{\alpha-1}}} \,,
\label{eqlossrate} 
\end{align} 
which is the solution to the differential equation
\begin{align} 
\frac{\dot{N}}{N_0} = -\frac{1}{\tau}\left(\frac{N}{N_0}\right)^\alpha\, ,
\end{align}
describing losses due to a few-body process of order $\alpha$. From this fit, applied to our data, we consistently obtain values of $\alpha$ close to 2, indicating that the losses are well described by two-body loss processes, and compute $L_2 = \tau^{-1}\frac{ V}{N_0}$ where $V = 2\sqrt{2}\pc{ \frac{2\pi k_B T}{\bar \omega ^2 m }}^{3/2}$ is the effective volume of the thermal sample, with $k_B$ the Boltzmann constant, $\bar \omega$ the geometric mean of the trapping frequencies, $T$ the gas temperature and $m$ the atomic mass.

To verify the robustness of our analysis, we followed an approach similar to Ref.\cite{Lecomte2024}. In this method, we prepare non-degenerate atomic samples at the same initial temperature but with varying total atom numbers. We then monitor the time evolution of the sample and extract the initial loss rate from a linear fit near the origin (see Fig.\ref{figL2Linear}a). 
By fitting the loss rate as a function of density, we confirm that it scales linearly, which is consistent with two-body loss processes (see Fig.~\ref{figL2Linear}b). This procedure was repeated at several different magnetic fields, and in each case, we found excellent agreement between this calibration of the two-body loss rate and the one obtained by fitting Eq.~\eqref{eqlossrate} to the data with $\alpha=2$.

\end{document}